# Electrodynamics of surface-enhanced Raman scattering

E. J. Adles

Department of Physics, NC State University Raleigh, NC 27695-8202

Applied Physics Laboratory, Johns Hopkins University, Laurel MD 20723

S. Franzen

Department of Chemistry, NC State University Raleigh, NC 27695-8204

Department of Chemistry, Zhejiang University, Hangzhou, China

D. E. Aspnes

Department of Physics, KyungHee University,

Seoul 130-701, KOREA; Department of Physics,

NC State University Raleigh, NC 27695-8202

(6 Sep 2011)

## Abstract

We examine SERS from two perspectives: as a phenomenon described by the Laplace Equation (the electrostatic or Rayleigh limit) and by the Helmholtz Equation (electrodynamic or Mie limit). We formulate the problem in terms of the scalar potential, which simplifies calculations without introducing approximations. Because scattering is not usually calculated this way, we provide the necessary theoretical justification showing that the scalar-potential description is complete. Additional simplifications result from treating the scatterer as a point charge $q$ instead of a dipole. This allows us to determine the consequences of including the longitudinal (Coulomb) interaction between $q$ and a passive resonator. This interaction suppresses the mathematical singularities that lead to the unphysical resonant infinities in first and second enhancements. It also modifies the effective restoring-force constant of a resonant denominator, which permits us to explore the possibility of dual resonance through a molecular pathway. We apply the formalism to spherical inclusions of radius $a$ for $q$ located at polar and equatorial positions. For small $a$, of the order of 1 nm or less, the low-$l$ multipole terms are important. For the more relevant case of radii of the order of 10 nm and larger, the $q$-sphere interaction can be approximated by a model where $q$ interacts with its image charge for a dielectric plane, and the singularity shifts in a discrete manner from $\varepsilon(\omega) = -2$ to $\varepsilon(\omega) = -1$. These results are supported by more accurate calculations taking retardation into account, although the use of only one spherical-harmonic term does not fully represent the difference between forward- and backscattering.

I. INTRODUCTION

Surface-enhanced Raman scattering (SERS), discovered over 30 years ago, [1, 2] has experienced renewed popularity as a result of the present interest in nanostructures and reported observations of SERS from single molecules. [3,4] A surprising aspect is the apparently extreme amplification of Raman intensities, with factors of the order of $10^{14}$ being claimed. [3,4] These factors are well beyond those that can be realized by enhancing the driving field by plasmonic resonances. [5–8] Additional mechanisms appear to be required to explain the reported enhancement factors. [9–11]

The concept of "hot spots" has been introduced to account for possible concentrations of the field in specific regions between dimer/multimer particles, sufficiently curved surfaces, or self-similar fractal aggregates. [10, 12–15] Of concern is that the average enhancement in a sample should be measurable as a bulk dielectric-response effect if the "hot spots" comprise a significant part of the sample, [16,17] as modeled for example by effective-medium theory. [18,19] An alternative view is that the electromagnetic enhancement factors are in fact modest, and that other mechanisms give rise to the observed scattering. [17] In this case, the molecular resonance may be important in a separate mechanism, such as resonance Raman. In parallel work, we have used the time-correlator formalism for calculation of resonance Raman cross sections [20] to show that molecular resonances are the most likely origin of the so-called chemical mechanism of SERS. [21–23] Yet another view has been presented that treats the local field, molecular resonance and charge transfer interactions in a Herzberg-Teller or vibronic coupling approach. [24]

Almost every treatment of this problem concedes that multiple mechanisms are required in order to achieve an enhancement factor of $10^{14}$. Although the electromagnetic mechanism is considered to be the dominant mechanism by many researchers, essentially every theory also includes the chemical mechanism. The nature of the chemical mechanism itself is also a subject of controversy, in part because the methods required to calculate resonance Raman cross sections have also been under development. We have advanced the view that explicit calculation of excited state displacements is essential for calculation of resonance Raman scattering. [21–23] The alternative to explicit calculation of the excited state displacements is the use of polarizability derivatives to calculate the Raman cross section. [25–27] However, the polarizability derivative approach is only valid far from resonance. Therefore, polarizability derivatives cannot accurately estimate the role played by the resonance Raman effect in the chemical mechanism. Explicit calculation of excited-state displacements has shown that pyridine resonance can be enhanced by optical excitation of Pyr-Ag clusters of up to 20 Ag atoms. [23] The role of excited-state displacements has been verified experimentally to be consistent with resonance Raman scattering from molecule-cluster supermolecules. [28] Regardless of the method used to calculate its magnitude, the chemical mechanism arises from a direct, i.e. bonding, interaction of the metal cluster with the adsorbate that leads to greater Raman enhancement. The electromagnetic theory can be interpreted as an enhancement of molecular polarizability that arises from the polarization of the inclusion by the incident radiation. [7,8] Since the molecule is close to a curved metal surface, a self-consistent solution of Maxwell's Equations leads to a local-field contribution that can be cast into an effective polarizability for the bound adsorbate. [29–31] The polarizability tensor couples the incident and scattered fields. Theories of electromagnetic enhancement have required enhancement in both of these fields to explain the reported giant enhancements. To simplify the physics, investigators have examined the incident and scattered field enhancements separately, and given the names first and second enhancement, respectively.

Electromagnetic, or more accurately, electrostatic, theory is invoked to explain observed large enhancements as the contribution of the outgoing Raman-scattered radiation by a second plasmonic resonance.[6, 32] Since the intensity of the incident and scattered radiation are each proportional to the square of the field, dual plasmonic resonances would allow large overall gains to be achieved with electric-field enhancements of the order of $10^3$ per stage. However, this argument fails when bandwidth is taken into account. To resonantly enhance two signals at somewhat different frequencies, it is necessary to trade Q for bandwidth. This places an upper limit to possible enhancements by purely plasmonic resonances. This aspect has been ignored in prior work with two exceptions. [33, 34] Yet it is evident that large enhancements are possible only if the quality factor $Q$ of the resonance is extremely high, which requires the resonance itself to be extremely narrow. [33] Almost every prior study has ignored damping so that the resonance was treated in the limit of an infinitely narrow and infinitely large enhancement. However, infinities are nonphysical. In a proper formulation of resonance enhancement they should not occur, and their appearance indicates that the elementary theory is deficient in some way. We show that the deficiency results from neglecting the longitudinal (Coulomb) interaction between the charge $q$ and the passive resonator. The inclusion modifies the response of $q$ so that at a resonance its motion is effectively frozen. While our calculations show that this occurs under conditions too extreme to have any practical effect, they identify and eliminate an objection to the elementary approach. We also provide an explicit general formulation of the enhancement problem in the electrostatic limit, then use it for spherical resonators to demonstrate our conclusions regarding intrinsic limits to first and second enhancements. As a consequence only one of the interacting electromagnetic waves, incident or scattered, can be at maximum gain. These calculations show that for molecules on a single isolated nanoparticle (inclusion), electromagnetic enhancement is limited to $10^6$ on Ag and significantly less on Au.

Our approach takes advantage of a point-charge model, which in anisotropic-bond form provided new insights into the physics of second-harmonic generation and other nonlinear-optical processes. [35] We start with a configuration $C$ and subdivide it into $q$ and a remaining system $S$, which we take to be a passive resonator of dielectric function $\varepsilon(\omega)$ embedded in a dielectric matrix. We calculate the radiation emitted by $C$ by evaluating the four basic steps that occur whenever a passive illuminated configuration radiates light: (1) determine the local field at $q$; (2) solve the force equation to obtain the acceleration of $q$; (3) calculate the resulting far-field radiation emitted by $q$, and (4) superpose the radiation from all $q$. In the present application, step (3) is expanded to include the radiation emitted by $S$ as well as $q$, and step (4) can be ignored. This general approach was originally applied to linear optics by Ewald, [36] and Oseen, [37] and is formally related to the derivation of the Lorentz polarizability as an optical response. It is not intended to replace a full quantum-mechanical treatment, but to elucidate the underlying physics.

As shown by numerous examples, [38–42] this approach is particularly effective when the emerging radiation is orders of magnitude weaker than the driving radiation and occurs at a different frequency, as is the case here. Under these conditions all four steps are independent, so the self-consistency difficulties that complicate linear-optics applications do not arise. However, the Raman-scattering problem is complicated by the presence of the passive resonator, which changes the resonant conditions of $q$ even though it nominally amplifies incoming fields through plasmon resonances and serves as a secondary radiator at the Raman-shifted frequencies. While the ratio of field strengths can be described by the reciprocity theorem, as recently noted by Le Ru and Etchegoin, [43] this is irrelevant because it does not take into account the reduction in emitted radiation caused by the connection between *q* and the resonator. A self-consistent solution is required, although this solution occurs at a different level from that of Ewald and Oseen.

Our approach of casting the calculation in terms of a bond charge instead of the usual point dipole is a seemingly minor detail, but it significantly reduces mathematical complexity. By selectively postponing differentiation until the final step, we avoid the need to use two arbitrarily large Green functions of opposite sign with infinitesimally small separation between singularities. This allows us to obtain for example the solution of a Raman-active molecule at an equatorial site on a spherical inclusion, which is relevant for aromatic molecules that are expected to lie flat on a surface. We find that the polarization of the resonator by *q*, essentially a self-consistent mass-renormalization interaction from the perspective of *q*, is present at all frequencies, driving as well as radiating, and can act to move native molecular excitations on or off resonance. Thus for single inclusions our calculations reveal four electromagnetic routes to enhancement: plasmon resonances in the passive resonator for both incoming and outgoing waves, and resonant enhancement of the associated electronic transitions at the incoming and outgoing frequencies in the Raman-active molecule. However, finite-bandwidth effects make it obvious that all cannot be active at once. [33]

Taken together, our results show that the extreme enhancements reported, if real, cannot be achieved with single particles but must involve a combination of the mechanisms discussed above in configurations involving two or more interacting particles. [15, 44] We shall discuss this case elsewhere. The net result is a detailed understanding of the electromagnetic enhancement mechanisms of SERS.

## II. THEORY

### A. Preliminaries

In this section we justify our use of the scalar potential $\phi$ as a sufficient descriptor in the current development. We begin by noting that electromagnetic scattering is described by the Helmholtz equation. For source-free regions and a harmonic time dependence $e^{-i\omega t}$, it reduces to

$$(\nabla^2 + \vec{k}^2)\psi = 0, \tag{2.1}$$

where $\psi$ is the scalar potential $\phi$, or a component of the electric field $\vec{E}$, the magnetic field $\vec{H} = \vec{B}$, or the vector potential $\vec{A}$. When sources are present, the solution of the inhomogenous Helmholtz equation can be written in 4-vector form as

$$(\phi, \vec{A}) = e^{-i\omega t} \frac{1}{c} \int d^3 r' (c\rho(\vec{r}'), \vec{J}(\vec{r}')) G(\vec{r}, \vec{r}'), \tag{2.2}$$

where $\rho$ and $\vec{J}$ are charge and current densities, respectively, and for empty space

$$G(\vec{r}, \vec{r}') = \frac{e^{ik|\vec{r} - \vec{r}'|}}{|\vec{r} - \vec{r}'|} \tag{2.3}$$

is the Green function. For scattering from spherical inclusions (Mie theory) the incoming plane wave is expressed in terms of various operations on spherical Bessel functions, and the outgoing wave in Hankel functions $h_j^{(1)}$. With these functions it is not necessary to divide the treatment into local and far-field parts, but these functions do not admit to simple treatments of the $q$-$S$ interaction so we continue developing the scalar potential.

For environments that are sufficiently local so retardation effects do not matter, we let $\vec{k} \to 0$ in Eq. (2.3) and recover the electro- and magnetostatic expressions

$$(\phi, \vec{A}) = e^{-i\omega t} \frac{1}{c} \int d^3 r' \frac{(c\rho(\vec{r}'), \vec{J}(\vec{r}'))}{|\vec{r} - \vec{r}'|}, \tag{2.4}$$

where $1/|\vec{r} - \vec{r}'|$ is the Green function of the Laplace Equation. For $|\vec{r}| > |\vec{r}'|$ the scalar potential can be written

$$\phi(\vec{r}) = \frac{1}{r} \int d^3 r' \rho(\vec{r}) + \frac{1}{r^2} \hat{r} \cdot \int d^3 r' \vec{r}' \rho(\vec{r}) + \ldots \tag{2.5}$$

where the dipole

$$\vec{p} = \int d^3 r' \vec{r}' \rho(\vec{r}) \tag{2.6}$$

can be identified as the vector part of the term proportional to $1/r^2$ in the expansion of the local scalar potential. In the radiation zone $|\vec{r}| \to \infty$, so we ignore the term $\vec{r}'$ in the denominator and expand the exponential to obtain

$$\phi(\vec{r},t) = -i\frac{e^{ikr-i\omega t}}{rc}\vec{k}\cdot\int d^3r'\vec{r}'\rho(\vec{r}') = -i\frac{e^{ikr-i\omega t}}{rc}\vec{k}\cdot\vec{p}. \tag{2.7}$$

Thus the dipole obtained from the $r^{-2}$ term in the expansion of the local potential is the appropriate dipole for $\phi(\vec{r},t)$ in the radiation zone. (2.8)

We still need to show that it is sufficient to work only with $\phi$. From the same expansion

$$(\phi,\vec{A}) = \frac{e^{ikr-i\omega t}}{rc}\int d^3r' e^{-i\vec{k}\cdot\vec{r}'}(c\rho(\vec{r}'),\vec{J}(\vec{r}')). \tag{2.9}$$

With the $r$ dependence appearing only in the prefactor, we apply the definition of the Lorentz gauge

$$\nabla\cdot\vec{A} + \frac{1}{c}\frac{\partial\phi}{\partial t} = 0 \tag{2.10}$$

and obtain $\phi = \hat{k}\cdot\vec{A}$. That is, $\phi$ is simply the longitudinal component of $\vec{A}$. Since $\rho$ and $\vec{J}$ are related by charge conservation, the other two components of $\vec{A}$ are connected to $\phi$ as well. Hence we can work with either $\phi$ or $\vec{A}$. The electric field $\vec{E}^{ff}$ in the far-field zone is then

$$\vec{E}^{ff}(\vec{r},t) = -\frac{1}{c}\frac{\partial\vec{A}}{\partial t} - \nabla\phi = ik(\tilde{I} - \hat{k}\hat{k})\cdot\vec{A}. \tag{2.11}$$

As the final step, standard treatments employ boundary conditions on $\vec{E}$ and $\vec{H}$. However, in the present application $\vec{A}$ is continuous everywhere, so the boundary conditions on $\phi$ are sufficient.

**B. General electrostatic formulation.**

As noted above, we consider a configuration $C$ consisting of a Raman-active molecule and a passive resonator, where the resonator is a polarizable inclusion of dielectric function $\varepsilon = \varepsilon(\omega)$ that is embedded in a uniform background medium of dielectric function $\varepsilon_a = 1$. Setting $\varepsilon_a = 1$ simplifies the equations with no loss of generality, since $\varepsilon$ is always normalized to $\varepsilon_a$. We let the relevant part of the molecule be a point charge $q$, which may be an element of a continuum charge density. We define the system $S$ to be everything but $q$, although in the present development the only other system element is the passive resonator. The response of $C$ to an externally applied field $\vec{E}_o(\vec{r},t) = \vec{E}_o(\vec{r})e^{-i\omega_o t}$ then consists of the responses of $S$ and $q$ to $\vec{E}_o(\vec{r},t)$, conditioned by their mutual interaction. Here, $\vec{r}$ is the location of the observer in a coordinate system where the origin is at the center of mass of the inclusion.

If the inclusion is near or at resonance at $\omega_o$, the field in the vicinity of the inclusion is larger than $\vec{E}_o(\vec{r},t)$ (first enhancement). For simple configurations the calculation of the enhancement is routine. Because we are focused here on the implications of the *q-S* interaction on other enhancement mechanisms, we assume that first enhancement is already pesent, and that the field in the vicinity of the resonator is

$$\vec{E}_1(\vec{r})e^{-i\omega_o t} = \vec{E}_o(\vec{r})e^{-i\omega_o t} + \vec{E}_s(\vec{r})e^{-i\omega_o t}, \tag{2.12a}$$

where $\vec{E}_s(\vec{r})e^{-i\omega_o t}$ is the enhancement due to *S*. We describe $\vec{E}_s(\vec{r})e^{-i\omega_o t}$ by the potential $\phi_s(\vec{r})e^{-i\omega_o t}$. For simplicity we assume that all boundaries are fixed, and ignore retardation and spatial dispersion. In this case the driving field for *q* is

$$\vec{E}_1(\vec{r})e^{-i\omega_o t} = \vec{E}_1 e^{-i\omega_o t}. \tag{2.12b}$$

Retardation is relevant for determining the upper limits to resonator sizes; we will discuss the special case of spherical resonators below.

When *q* is inserted at location $\vec{r}\,'$, new potentials that involve *q* and the associated polarization of *S* arise. We write the new overall potential as

$$\phi_{tot}(\vec{r},\vec{r}\,') = \phi_s(\vec{r}) + \phi_{sq}(\vec{r},\vec{r}\,') + \phi_q(\vec{r},\vec{r}\,'), \tag{2.12c}$$

where

$$\phi_q(\vec{r},\vec{r}\,') = \frac{q}{|\vec{r}-\vec{r}\,'|}, \tag{2.12d}$$

$\phi_{sq}(\vec{r},\vec{r}\,')$ is the contribution of the polarization induced in *S* by *q*, and a time dependence of $e^{-i\omega_o t}$ is assumed. By separating the overall potential into these three terms, we identify and hence can calculate the consequences of the interaction between *q* and *S*, and therefore the response of *C* to the external driving field. The respective fields associated with each potential are given as usual by their negative gradient with respect to $\vec{r}$.

An equilibrium position $\vec{r}_q\,'$ is established by balancing the forces exerted on *q* by the fields and an intrinsic restraint that we represent for now with a constant second-rank Hooke's-Law restoring-force tensor $\vec{\kappa}$. Because the driving field has the time dependence $e^{-i\omega_o t}$, all potentials considered so far will also have this time dependence. Because the restraint is not perfectly rigid, $q$ oscillates about $\vec{r}_q'$ with the same time dependence. Accordingly, to first order we write $\vec{r}' = \vec{r}_q' + \Delta \vec{r}_o' e^{-i\omega_o t}$, where $\Delta \vec{r}_o'$ is a constant. The subscript $o$ on $\Delta \vec{r}_o'$ denotes that this displacement occurs at frequency $\omega_o$. When Raman scattering is introduced, we will define additional displacements $\Delta \vec{r}_\pm'$ associated with the Raman frequencies $\omega_\pm = \omega_o \pm \omega_v$, where $\omega_v$ is a phonon frequency.

## C. Equation of motion and far-field radiation

To proceed further we determine $\Delta \vec{r}_o'$. From Newton's Law of Motion

$$m \frac{d^2 \vec{r}'(t)}{dt^2} = m \frac{d^2}{dt^2} \Delta \vec{r}_o' e^{-i\omega_o t} = -m\omega_o^2 \Delta \vec{r}_o' e^{-i\omega_o t}$$

$$= q\{\vec{E}_o e^{-i\omega_o t} - [\nabla_{\vec{r}} \varphi_s(\vec{r}) e^{-i\omega_o t}]|_{\vec{r} = \vec{r}_q' + \Delta \vec{r}_o' e^{-i\omega_o t}} - [\nabla_{\vec{r}} \varphi_{sq}(\vec{r}, \vec{r}_q' + \Delta \vec{r}_o' e^{-i\omega_o t})]|_{\vec{r} = \vec{r}_q' + \Delta \vec{r}_o' e^{-i\omega_o t}}\}$$

$$- \vec{\kappa} \cdot \Delta \vec{r}_o' e^{-i\omega_o t}. \tag{2.13}$$

The quantity in braces is the local field, completing step 1. If the restoring force is isotropic such that $\vec{\kappa} = \kappa \vec{I}$, where $\vec{I}$ is the unit second-rank tensor, then $\kappa$ is related to the classical polarizability $\alpha$ by the Lorentz model

$$\alpha = \frac{e^2}{\kappa}. \tag{2.14}$$

To solve Eq. (2.13) for $\Delta \vec{r}_o'$, we linearize it by expanding all terms to first order in $\Delta \vec{r}_o' e^{-i\omega_o t}$. When this is done the first-order expansion of the second term has the time dependence $e^{-i2\omega_o t}$. Since we are only interested in the linear response, this term is discarded. To evaluate the third term, we first take the gradient with respect to $\vec{r}$, then substitute $\vec{r} = \vec{r}_q' + \Delta \vec{r}_o' e^{-i\omega_o t}$ before expanding. The result is

$$-[\nabla_{\vec{r}} \phi_{sq}(\vec{r}, \vec{r}_q' + \Delta \vec{r}_o' e^{-i\omega_o t})]|_{\vec{r} = \vec{r}_q' + \Delta \vec{r}_o' e^{-i\omega_o t}}$$

$$= -\nabla_{\vec{r}} \phi_{sq}(\vec{r}, \vec{r}_q')|_{\vec{r} = \vec{r}_q'} - \Delta \vec{r}_o' \cdot \nabla_{\vec{r}_q'}[\nabla_{\vec{r}} \phi_{sq}(\vec{r}, \vec{r}_q')|_{\vec{r} = \vec{r}_q'}] e^{-i\omega_o t} \tag{2.15a}$$

$$= (1 + e^{-i\omega t}\Delta\vec{r}_o{}'\cdot\nabla_{\vec{r}_q'})\vec{E}^o_{sq}(\vec{r}_q'),\tag{2.15b}$$

where

$$\vec{E}^o_{sq}(\vec{r}_q') = -\nabla_{\vec{r}}\phi_{sq}(\vec{r},\vec{r}_q')_{\vec{r}=\vec{r}_q'}.\tag{2.15c}$$

$\vec{E}^o_{sq}(\vec{r}_q')$ is a function of $\omega$ through $\varepsilon(\omega)$; the superscript $o$ indicates that $\vec{E}^o_{sq}(\vec{r}_q')$ is evaluated at $\omega_o$. Combining everything, the equation of motion is

$$-m\omega^2\Delta\vec{r}_o{}' = q\vec{E}_1(\vec{r}_q') + q(\Delta\vec{r}_o{}'\cdot\nabla_{\vec{r}_q'})\vec{E}^o_{sq}(\vec{r}_q') - \vec{\kappa}\cdot\Delta\vec{r}_o{}',\tag{2.16a}$$

where the time dependence $e^{-i\omega_o t}$ is understood, and where

$$\vec{E}_1(\vec{r}_q') = \vec{E}_o - \nabla_{\vec{r}_q'}\phi_s(\vec{r}_q') = \vec{E}_o + \vec{E}_s(\vec{r}_q')\tag{2.16b}$$

is the local driving field of Eqs. (2.12a) and (2.12b) evaluated at $\vec{r} = \vec{r}_q'$. In the most general situation we obtain $\Delta\vec{r}_o{}'$ by performing the necessary tensor transformations to isolate it on one side of Eq. (2.16a).

We now introduce Raman scattering by making $\vec{\kappa}$ time-dependent according to

$$\vec{\kappa} \to \vec{\kappa}(t) = \vec{\kappa} + \Delta\vec{\kappa}\cos(\omega_v t),\tag{2.17}$$

where $\Delta\vec{\kappa}$ is a constant. According to the classical model, this accounts mathematically for the modulation of the polarizability by a phonon. The phonon distorts what would otherwise be a simple restoring force, generating sidebands at frequencies $\omega_\pm = \omega_o \pm \omega_v$, and corresponding displacements $\Delta\vec{r}_\pm{}'$ that oscillate with angular frequencies $\omega_\pm$. For mathematical simplicity we ignore damping, which moves the singularities off the real axis. As shown by Franzen, [33] the effect of damping on field enhancement is severe and in practice cannot be neglected. However, our objective is to identify and quantify singularities. If needed, it is straightforward to include damping. Following the same procedure as before, we find that the $\Delta\vec{r}_\pm$ are given by

$$-m\omega^2\Delta\vec{r}_\pm{}' = \Delta\vec{\kappa}\cdot\Delta\vec{r}_o{}' + q(\Delta\vec{r}_\pm{}'\cdot\nabla_{\vec{r}_q'})\vec{E}^\pm_{sq}(\vec{r}_q') - \vec{\kappa}\cdot\Delta\vec{r}_\pm{}'\tag{2.18}$$

The driving term for the Raman fields is proportional to $\Delta\vec{r}_o{}'$, as expected.

The forms of Eqs. (2.16a) and (2.18) show that the *S-q* interaction generates a correction to $\vec{\kappa}$, and that corrections will occur at all frequencies. As a result, pre-existing resonance frequencies within the molecule are shifted. If these shifts move the molecule into resonance, this provides another route to enhancement. However, since the pump and Raman frequencies are different, not all resonances will occur simultaneously. [33]

Calculation of the contribution to the far-field radiation from *q* is straightforward, because the functional form is known. For $\phi_{sq}(\vec{r},\vec{r}')$ the situation is complicated by the fact that we cannot solve $\nabla_{\vec{r}}^2 \phi_{sq}(\vec{r},\vec{r}') = 0$ for $\phi_{sq}(\vec{r},\vec{r}')$ unless we know the configuration. However, we can still draw some generally valid conclusions. First, the boundary conditions ensure that the introduction of *q* does not change the charge on *S*, so an initially neutral inclusion remains neutral. Second, we gain information about magnitudes by noting that the part of $\phi_{sq}(\vec{r},\vec{r}')$ that radiates is the change $\Delta\phi_{sq}(\vec{r},\vec{r}')$ that results from the motion of *q*. For a generic $\Delta\vec{r}'$ this is given by

$$\Delta\phi_{sq}(\vec{r},\vec{r}') = (\Delta\vec{r}'\cdot\nabla_{\vec{r}'})\phi_{sq}(\vec{r},\vec{r}') . \tag{2.19}$$

Upon expansion, this yields the radiating dipole $\Delta\vec{p}_{sq}$.

In principle arbitrarily large second-enhancements occur because in the absence of damping $\phi_{sq}(\vec{r},\vec{r}')$, and therefore $\Delta\phi_{sq}(\vec{r},\vec{r}')$ and $\Delta\vec{p}_{sq}$, can become arbitrarily large at resonance. However, Eqs. (2.5a) and (2.7) show that a corresponding divergence occurs in the prefactor to $\Delta\vec{r}'$, so an overall singularity is avoided. Because we reach this conclusion independent of any specific configuration, the result is general.

### III. APPLICATION

#### A. Spherical inclusions, *q-S* interaction.

To illustrate the above, we consider *S* to be a spherical inclusion of radius *a* and dielectric function $\varepsilon = \varepsilon(\omega)$. We take the incident beam to be polarized along *z*, and assume that all dimensions are small compared to the wavelengths $\lambda_o$ and $\lambda_z$ associated with $\omega_o$ and $\omega_z$, respectively. The electrostatics problem for a small sphere in a uniform field is standard, and the resulting potential for $|\vec{r}| > a$ includes only the driving field and the dipole term from the sphere:

$$-E_o r\cos\theta + \phi_s(\vec{r}) = -E_o r\cos\theta + E_o\left(\frac{\varepsilon_o - 1}{\varepsilon_o + 2}\right)\frac{a^3}{r^2}\cos\theta, \tag{3.1}$$

where the subscript $o$ on $\varepsilon_o$ means the $\varepsilon$ is evaluated at the driving frequency $\omega = \omega_o$. The plasmon enhancement due to the sphere and its singularity at $\varepsilon_o = -2$ are evident. A primary characteristic of the above is the dependence of the magnitude of first enhancement on the sphere volume, as encoded in $a^3$. This is a consequence of $\vec{A}$ being proportional to the integral of $\vec{J} = \frac{-i\omega(\varepsilon_o - 1)}{4\pi}\vec{E}$, which is constant within the sphere. In practice this is reduced to a linear dependence, because the molecule can never get closer to the sphere than the radius $a$.

We next consider the interaction between $q$ and $S$. Calculation of the associated potential is again straightforward. With $q$ at $\vec{r}\,' = (r',\theta',\phi')$, the general solution for the potential outside $S$ at a frequency $\omega$ is

$$\phi_q(\vec{r},\vec{r}\,') + \phi_{sq}(\vec{r},\vec{r}\,') = q\sum_{l=0}^{\infty}\frac{4\pi}{2l+1}\frac{(r')^l}{r^{l+1}}\sum_{m=-l}^{l}Y_{lm}^*(\theta',\phi')Y_{lm}(\theta,\phi)$$

$$+ q\sum_{l=0}^{\infty}\frac{4\pi l}{2l+1}\frac{a^{2l+1}}{(rr')^{l+1}}\frac{\varepsilon - 1}{l(\varepsilon + 1) + 1}\sum_{m=-l}^{l}Y_{lm}^*(\theta',\phi')Y_{lm}(\theta,\phi), \tag{3.2}$$

where $r' < r$ in the first term and $a < r, r'$ in the second. The first term on the right is the series expansion of the potential due to $q$, and the second from the polarization that $q$ induces in the sphere. The sums over the index $m$ concern only the spherical harmonics, so we apply the sum rule

$$\sum_{m=-l}^{l}Y_{lm}^*(\theta',\phi')Y_{lm}(\theta,\phi) = \frac{2l+1}{4\pi}P_l(\cos\gamma) \tag{3.3a}$$

where $P_l(x)$ is a Legendre polynomial and

$$\cos\gamma = \sin\theta\sin\theta'\cos(\phi - \phi') + \cos\theta\cos\theta'. \tag{3.3b}$$

We obtain

$$\phi_q(\vec{r},\vec{r}\,') + \phi_{sq}(\vec{r},\vec{r}\,') = q\sum_{l=0}^{\infty}\left(\frac{(r')^l}{r^{l+1}} + \frac{l(\varepsilon - 1)}{l(\varepsilon + 1) + 1}\frac{a^{2l+1}}{(rr')^{l+1}}\right)P_l(\cos\gamma) \tag{3.3c}$$

The field due to the polarization of the sphere induced by $q$ is therefore

$$\vec{E}_{sq}(\vec{r},\vec{r}') = q \sum_{l=0}^{\infty} \frac{l(\varepsilon-1)}{l(\varepsilon+1)+1} \frac{a^{2l+1}}{r^{l+2}(r')^{l+1}}$$

$$\times \left( -\hat{r}(l+1) + \frac{dP(\cos\gamma)}{d(\cos\gamma)} [\hat{\theta}\sin(\theta'-\theta) - \hat{\phi}\sin\theta\sin\theta'\sin(\phi-\phi')] \right) \quad (3.4)$$

To calculate the back-reaction of $S$ on $q$ we set $\vec{r} = \vec{r}'$, in which case $\cos\gamma = P_l(\cos\gamma) = 1$. We obtain

$$\vec{E}_{sq}(\vec{r}') = -q\hat{r}' \sum_{l=0}^{\infty} \frac{l(l+1)(\varepsilon-1)}{l(\varepsilon+1)+1} \frac{a^{2l+1}}{(r')^{2l+3}} \quad (3.5)$$

The field at $q$ is entirely radial, consistent with a longitudinal electrostatic interaction with retardation effects neglected. Hence for the $q$-$S$ interaction the choice of coordinate system is irrelevant. Because $\vec{E}_{sq}(\vec{r}')$ is proportional to $\hat{r}'$ rather than to $\hat{z}$, lateral forces perpendicular to $z$ will be generated when $q$ is driven off-axis, relevant in the equatorial location discussed below. The infinite number of singularities in Eq. (3.5) at values

$$\varepsilon = \varepsilon(\omega) = \varepsilon_l = -1 - 1/l \quad (3.6)$$

represent the different plasmonic resonances to which $q$ couples.

Unless the sphere is very small, the ratio $a/r' = a/d$, where $d$ is the distance of $q$ from the center of the sphere, will be very close to 1. In this case the sum is dominated by high-$l$ terms. Assuming that these dominate completely, we ignore the 1 in the denominator and rewrite Eq. (3.5) as

$$\vec{E}_{sq}(\vec{r}') \approx -q\hat{r}' \left(\frac{\varepsilon-1}{\varepsilon+1}\right) \sum_{l=0}^{\infty} (l+1) \frac{a^{2l+1}}{(d)^{2l+3}} \approx -\frac{q\hat{r}'}{4(d-a)^2} \left(\frac{\varepsilon-1}{\varepsilon+1}\right). \quad (3.7)$$

To obtain the closed-form solution we assume that $d \approx a$, which is certainly valid for Raman-active molecules adsorbed onto spherical inclusions of nominally optimal enhancement sizes of ca. 50 nm.

We recognize Eq. (3.7) as the field at $q$ due to its image charge at $(-d)$ at a planar boundary of a semi-infinite medium of dielectric function $\varepsilon_o$. This is physically reasonable, because in the limit of large diameters there is little distinction between $q$ being adjacent to the sphere or to a semi-infinite plane. The function of the higher-order terms in the original expansion is also clarified: they narrow the lateral extent of the field to a width of the order of $(d-a)$. A main point is that there exists a countably infinite set of resonances from $\varepsilon = -2$ to $\varepsilon = -1$ corresponding to the different multipoles as the sphere diameter increases. In principle any of these singularities forces the displacement of $q$ to zero.

## B. Spherical inclusion, first enhancement.

Although a solution can be obtained for any location of the Raman-active molecule, for illustrative purposes we consider the polar and equatorial-plane positions, with the molecule on the $z$ axis at $z = d > a$ in the former case and on the $x$ axis at $x = d > a$ in the latter case. The local field of the sphere is twice as large at a pole, but the equatorial location is more appropriate for Raman-active molecules that are physi- or chemisorbed to the sphere since many Raman-active dye molecules adsorb to metal inclusions such that their transition moments are parallel, rather than perpendicular to the surface. The negative gradient with respect to $\vec{r}$ of Eq. (3.1) gives

$$E_1(\hat{z}r) = \hat{z}E_o\left[1 + 2\left(\frac{\varepsilon_o - 1}{\varepsilon_o + 2}\right)\frac{a^3}{r^3}\right] \tag{3.8a}$$

at the $\theta = 0$ pole and

$$E_1(\hat{x}r) = \hat{z}E_o\left[1 - \left(\frac{\varepsilon_o - 1}{\varepsilon_o + 2}\right)\frac{a^3}{r^3}\right] \tag{3.8b}$$

at the equator. Again, the subscript o designates evaluation at the pump frequency $\omega = \omega_o$. Because $r$ can never be less than $a$, the maximum fields are obtained with the molecule as close as possible to the sphere. The amplification is purely plasmonic, and independent of the size of the sphere. Although in the equatorial case $q$ moves off the equatorial plane, we can use the zero-order driving fields because small off-axis motions generate harmonics that are not relevant to the current problem. Also, we assume that the local field at any region of the molecule is the same as that at its center, which is acceptable since the Raman response is already first order in small quantities. Hence we can use Eqs. (3.8) as-written.

## C. Spherical inclusion, Second enhancement.

The role of the second enhancement at the Raman shifted frequency is central to the electromagnetic mechanism of SERS. To investigate this enhancement, we consider the polar and equatorial locations as above. The response at each location is divided into the charge motion and Raman radiation by the adsorbate.

### C.1 Charge motion, polar location

For a molecule adsorbed at a pole, $\vec{r}_q' = \hat{z}d$ and $\Delta\vec{r}_q' = \hat{z}\Delta d$. Thus motion is strictly along the $z$ axis. If $\kappa$ is isotropic, Eq. (2.15a) for the response $\Delta d_o$ of $q$ to the driving field becomes

$$-m\omega^2 \Delta d_o = qE_1 + qE_{sq}^{o\prime} \Delta d_o - \kappa \Delta d_o, \tag{3.9a}$$

where

$$qE_1 = qE_o\left(1 + 2\frac{\varepsilon_o - 1}{\varepsilon_o + 2}\frac{a^3}{d^3}\right) \tag{3.9b}$$

and the effective spring constant $qE_{sq}^{o\prime}$ of the $q$-$S$ interaction is

$$q\vec{E}_{sq}^{o\prime} = q^2\left(\sum_{l=0}^{\infty} \frac{l(l+1)(2l+3)(\varepsilon_o - 1)}{l(\varepsilon_o + 1) + 1} \frac{a^{2l+1}}{d^{2l+4}}\right) \tag{3.9c}$$

$$= \frac{10q^2 a^3}{d^6}\left(\frac{\varepsilon_o - 1}{\varepsilon_o + 2}\right) \tag{3.9d}$$

$$= \frac{q^2}{2(d-a)^3}\left(\frac{\varepsilon_o - 1}{\varepsilon_o + 1}\right) \tag{3.9e}$$

where Eqs. (3.9c), (3.9d), and (3.9e) refer to the general, resonant $l = 1$, and large-sphere conditions, respectively, in the equation of motion

$$\Delta d_o = \frac{qE_1}{\kappa - m\omega^2 - qE_{sq}^{o\prime}}. \tag{3.10}$$

Assuming no broadening, at $\varepsilon_o = -2$ the singularity in the numerator is cancelled by that of Eq. (3.9d), capping $\Delta d_o$ at

$$q\Delta d_o |_{max} = E_o d^3 / 5. \tag{3.11}$$

The increase as the third power of the distance from the center of the sphere is understood in physical terms because at resonance for large $d$ the $q$-$S$ interaction effectively changes from a $q$-plane interaction at small spacing to a $q$-dipole interaction with the dipole at the center of the sphere.

Before proceeding into a similar discussion on the Raman-radiation part of the problem, we consider some numbers. We suppose a 1 watt pump beam at $\lambda = 514.5$ nm focused onto a 10 μm spot. Then $\omega = 3.66\times 10^{15}$ s$^{-1}$, and if we take $m = m_e = 9.11\times 10^{-31}$ kg, then $m\omega^2 = 14.4$ N/m establishes a benchmark for the remaining terms. Assuming in addition a ≈ d = 25 nm and (d – a) = 0.2 nm, the effective spring constant of the $q$-$S$ interaction is

$$q\vec{E}^o_{sq}{}' = 14.4 \frac{N}{m}\left(\frac{\varepsilon - 1}{\varepsilon + 1}\right), \qquad (3.12a)$$

while the $l = 1$ term only evaluates to

$$q\vec{E}^o_{sq}{}' = (1.5 \times 10^{-4} \frac{N}{m})\left(\frac{\varepsilon - 1}{\varepsilon + 2}\right). \qquad (3.12b)$$

This 5-order-of-magnitude difference in prefactors indicates that lifetime broadening contributions make the resonant term of academic interest only, and that the major effect of the $q$-$S$ interaction for nanospheres of the order of 10 nm radius is not only to inhibit the motion of the charge, but also to significantly shift the internal resonances of the molecule.

### C.2 Raman radiation, polar location

To calculate the Raman radiation we replace the drive term $qE_1$ in Eq. (3.9a) with $\Delta\kappa\Delta d_o$ and repeat the entire calculation for $\omega = \omega_\pm$, keeping $\varepsilon = \varepsilon_o$ in Eq. (3.9b) but using $\varepsilon = \varepsilon_\pm$ in Eqs. (3.9c-e). The singularities in the drive and Raman terms occur at different frequencies. The dipole for determining the far-field radiation is, from Eqs. (2.5), (2.6), and (2.18),

$$\Delta p = q\Delta d_\pm - 2\left(\frac{\varepsilon_\pm - 1}{\varepsilon_\pm + 2}\right)\left(\frac{a^3}{d^3}\right)(q\Delta d_\pm). \qquad (3.13)$$

This exhibits a new singularity at the Raman frequency. Although this singularity is technically cancelled by that in the denominator of Eq. (3.10), in practice the real amplitude limit is imposed by lifetime broadening. Thus the dominant effect of the $q$-$S$ interaction at the Raman-emission stage remains to modify the frequency at which potential molecular resonances occur. Hence the Raman-shifted signal can be enhanced, although a direct application of the reciprocity theorem without taking the $q$-$S$ interaction into account, as proposed by Etchegoin, would predict an incorrect frequency.

### C.3 Charge motion, equatorial location

We retain $\vec{E}_1$ along $z$ and locate $q$ on the $x$ axis at $r' = d$, $\theta = \pi/2$. The driving field is given by Eqs. (2.12a) and (2.12b). The motion $\Delta d_o$ is still in the $z$ direction, but $d$ remains constant and the gradient is taken with respect to the direction of $\vec{r}\,'$. The $q$-$S$ interaction term in the force equation is then

$$q\vec{E}^o_{sq}{'} = q^2 \left( \sum_{l=0}^{\infty} \frac{l(l+1)(\varepsilon_o - 1)}{l(\varepsilon_o + 1) + 1} \frac{a^{2l+1}}{d^{2l+3}} \right) \frac{1}{d}; \qquad (3.14a)$$

$$= \frac{q^2 a^3}{d^6} \left( \frac{\varepsilon - 1}{\varepsilon + 2} \right); \qquad (3.14b)$$

$$= \frac{q^2}{4a(d-a)^2} \left( \frac{\varepsilon - 1}{\varepsilon + 1} \right) \qquad (3.14c)$$

where Eqs. (3.14a), (3.14b), and (3.14c) refer to the exact, resonant, and large-sphere versions, respectively. The dielectric functions are to be evaluated at $\omega = \omega_o$ for the pump frequency and $\omega = \omega_\pm$ for the Raman radiation. It is seen that this term vanishes completely in the limit of large radius, as it must. In fact, using the previous numbers, the prefactor of the equatorial term is 0.058 N/m, which can be compared to 14.4 N/m found above the polar term. This result illustrates the substantial weakening of the restoring force with increasing sphere size relative to the situation for polar motion, where after a certain point sphere size becomes irrelevant. The prefactor for the resonant term is similarly reduced, to $-3.0 \times 10^{-5}$ N/m, although the prefactor is still mathematically necessary to eliminate the singularity in $\Delta d_o$ that would result if it were not present.

**C.4  Raman radiation, equatorial location**

The same discussion applies to the calculation of the Raman component of the displacement, with $\varepsilon(\omega)$ being evaluated at $\omega = \omega_\pm$. The $S$-$q$ radiation term is similarly reduced. The expressions for the dipoles are now

$$\Delta p = q\Delta d_\pm - \left( \frac{\varepsilon_\pm - 1}{\varepsilon_\pm + 2} \right)\left( \frac{a^3}{d^3} \right)(q\Delta d_\pm), \qquad (3.15)$$

so the apparent singularity at $\varepsilon_\pm = -2$ remains. However, as noted above, the expression remains finite because as $\varepsilon_\pm \to -2$, $\Delta d_\pm \to 0$ and the product remains finite. The capped value is now

$$\Delta p = \Delta \kappa \Delta d_o \frac{d^3}{2q}. \tag{3.16}$$

which is larger than the value given in Eq. (3.13) because the *S-q* restoring-force interaction is weaker.

## IV. DISCUSSION

### A. Finite-size effects

As noted in Sec. II, exact solutions of plasmonic scattering by a passive spherical resonator of any size are available as series expansions. These describe Mie scattering as a straightforward application of vector spherical harmonics. However, when a driving charge $q$ is present, these calculations becomes significantly more cumbersome. Kerker et al. have thoroughly investigated multiple-scattering effects, concluding that they are far below the level of detectability. However, they do not consider the longitudinal interaction. Their approach and its limitations can be summarized as follows.

First, the vector spherical harmonics are defined as [45]

$$\vec{X}_{lm} = \frac{1}{\sqrt{l(l+1)}} \vec{L} Y_{lm}(\theta, \phi) \tag{4.1}$$

where

$$\vec{L} = -i\vec{r} \times \nabla. \tag{4.2}$$

$\vec{L}$ is defined most conveniently in terms of raising and lowering operators

$$L_\pm = L_x \pm iL_y \tag{4.3}$$

where

$$L_+ Y_{lm} = \sqrt{(l-m)(l+m+1)}\, Y_{l,m+1},\quad L_+ Y_{lm} = \sqrt{(l+m)(l-m+1)}\, Y_{l,m-1} \tag{4.4}$$

and where

$$L_z Y_{lm} = m Y_{lm}. \tag{4.5}$$

In terms of spherical Bessel functions, the Green function for the Helmholtz equation is

$$\frac{e^{ik|\vec{r}-\vec{r}'|}}{|\vec{r}-\vec{r}'|} = 4\pi i k \sum_{l=0}^{\infty} j_l(kr_<)h_l^{(1)}(kr_>) \sum_{m=-l}^{l} Y_{lm}(\theta,\phi)Y_{lm}^*(\theta',\phi'),  \tag{4.6}$$

the expansion of a vector plane wave is,

$$(\hat{x}+i\hat{y})e^{ikz} = \sum_{l=1}^{\infty} i^l \sqrt{4\pi(2l+1)} \left[ j_l(kr)\vec{X}_{l1} + \frac{1}{k}\nabla \times \left(j_l(kr)\vec{X}_{l1}\right) \right], \tag{4.7}$$

and the potential of a point charge $q$ at $\vec{r}'$ is a scaled version of Eq. (4.6):

$$\phi_q = \frac{q}{4\pi} \frac{e^{ik|\vec{r}-\vec{r}'|}}{|\vec{r}-\vec{r}'|}. \tag{4.8}$$

Equation (4.6) can be verified by integrating it over all space, noting that the delta function for the angular coordinates is

$$\sum_{m=-l}^{l} Y_{lm}(\theta,\phi)Y_{lm}^*(\theta',\phi') = \delta(\cos\theta - \cos\theta')(\phi - \phi'). \tag{4.9}$$

To determine the vector plane wave in spherical Bessel functions we use Eq. (4.7), looking for a combination of functions that approach a constant value as $r \to 0$ as indicated on the left side of the equation. The first (magnetic-dipole) term contributes nothing in this case, but the necessary functions follow from the curl (electric dipole) term by noting in addition that

$j_l'(kr) = \lim_{r \to 0}\left(\frac{1}{kr} j_l(kr)\right)$. The important point is that the function that most closely approximates the actual behavior of the potential of the incoming field is

$$\phi_{incoming}(\vec{r}) = E_o \frac{1}{k} j_1(kr), \tag{4.10}$$

where the necessary angular functions are a consequence of taking the derivative with respect to any of the Cartesian coordinates. Choosing the finite and outgoing forms for the interior and exterior expansions we have

$$\phi_{sphere}(r) = A_1 j_1(k_1 r), \text{ and } \phi_{scatt}(\vec{r}) = B_1 h_1^{(1)}(k_2 r), \tag{4.11}$$

where $(ck_1/\omega)^2 = \varepsilon$ and $(ck_2/\omega)^2 = 1$ Matching boundary conditions at $r = a$ leads to

$$\vec{B}_l = \frac{k_2 a j_l(k_1 a) j_l'(k_2 a) - k_1 a \varepsilon\, j_l'(k_1 a) j_l(k_2 a)}{k_1 a \varepsilon\, j_l'(k_1 a) h_l^{(1)}(k_2 a) - k_2 a j_l(k_1 a) h_l^{(1)'}(k_2 a)} 3ik_2 q h_l^{(1)}(k_2 d) \tag{4.12a}$$

$$\approx -i \frac{\varepsilon - 1 + (k_1 a)^2 (1 - 3\varepsilon)/10 + (k_2 a)^2 (3 - \varepsilon)/10}{\varepsilon + 2 - (k_1 a)^2 (3\varepsilon + 2)/10 - (k_2 a)^2 (\varepsilon/2)} k_2^2 aq \frac{a^2}{d^2} (1 + (k_2 d)^2/2) . \tag{4.12b}$$

Thus the singularity near $\varepsilon \to -2$ remains, but the first consequence of finite size is to shift the resonance rather than decrease the signal. In fact whether the signal increases or decreases depends critically on the value of $\varepsilon$. With $\varepsilon \approx -2$ the coefficient actually increases with increasing sphere size. From the above expansion, if we pick $k_1 a \sim 0.3$ as a measure of where finite-size effects might start to become important, we find $a \sim 0.3 \lambda/(2\pi\sqrt{2}) \sim 17$ nm for 514.5 nm light. This is a surprisingly small value, but lends credence to the observations that enhancement is optimized in the 50 to 100 nm particle-size range.

**B. Radiation damping**

The above equations do not take radiation damping into account, although it is easily included by adding a viscous-friction term in the force equation. In principle the presence of the passive resonator improves the coupling of the radiating molecule to the outside world, with the molecule transferring energy to the passive resonator. However, this energy must come from the molecule. In the absence of dissipation the energy is traded between resonator and molecule as

$$W = (\Delta\kappa \Delta z_o') \Delta z_\pm' \tag{4.13}$$

As dissipation increases, the phase angle between $\Delta z_\pm'$ and $\Delta\kappa \Delta z_o'$ increases. However, there probably exists a maximum power level that the molecule can deliver. Inclusion of radiation damping has already been shown to lead to a finite bandwidth that defines the quality factor $Q$. The most significant aspect of radiation damping is to put further constraints on the magnitude of enhancements, which are only upper bounds in the model here.

**C. Alternative configurations**

Many different configurations have been proposed in attempts to explain huge enhancement factors. In principle all can be treated by the general theory presented here, which requires only $q$ and a system with which it interacts. Hence it applies to inclusions with depolarization factors other than 1/3, or for that matter a random collection of small particles. In a random collection there will probably always be a few for which resonant conditions are achieved. Note that the theory doesn't even require inclusions – it also describes random networks or contacting particles or whatever, as long as $q$ can induce polarization in its surroundings.

**D. Relationship to previous studies of the magnitude of the SERS effect**

All previous publications recognize plasmonic amplication at $\omega_o$ as a primary factor in SERS. Surprisingly, none of them recognize the need for self-consistency at $\omega_o$ although the need for self-consistency at the sideband frequencies is discussed in several papers. Even ignoring $\Delta\kappa$ and the $\omega_o$ issue, none of the previous works considered all of the remaining paths.

Gersten et al. [7, 8] and Metiu et al. [29, 30] considered in detail stage-1 amplification and self-consistency, with Gersten et al. recognizing explicitly that both could lead to enhanced emission (the large numerator and small denominator contributions, respectively, in the Gersten et al. terminology). However, neither recognized the possibility of second-stage amplification of Raman radiation by the sphere. In the Metiu et al. treatment this is surprising, because these authors based their work on a dyadic Green function that in principle contains the contribution of the sphere to the radiated field as well as treating the entire problem self-consistently. [29, 30] Although their result clearly exhibits the effect of self-consistency, second-stage amplification does not appear. Since Metiu et al. do not provide the Green function on which their work is based, it is not clear whether this absence is due to the use of an improper starting expression or an error in its reduction.

In the other major early contribution, Kerker et al. performed a thorough analysis of the combination of sphere and charge. [6] Although they ultimately reduced their results to the electrostatic limit, their work begins with the Green function of the time-dependent Helmholtz equation, meaning that it is valid for any size sphere and any separation between charge and sphere. However, in their formalism they do not consider self-consistency the polarizability of sphere and the intrinsic molecular polarizability in the absence of the sphere. Thus the small-denominator contribution is missing in their work. A more recent attempt to include self-consistency in the treatment of the effective polarizability by Masiello and Schatz is valid only at $\omega_o$, and hence only treats Rayleigh scattering not Raman scattering. [31]

**V. CONCLUSION**

We have used the anisotropic bond model to guide the development of a general theory of the electromagnetic aspects of SERS. The results show that for each Raman sideband there are four electromagnetic routes to enhancement, plasmonic amplification and resonance absorption/ emission in the molecule at both driving $\omega_o$ and Raman $\omega_\pm$ frequencies. However, for quality factors $Q$ high enough to explain claimed enhancements, only one of each type of enhancement is possible. This is because the Raman shift is sufficiently large for any important vibrational mode that it is not possible for both the incident and scattered frequencies to be near the maximum of their respective enhancement functions. Moreover, unless there is a loss mechanism in the conductor there will be no energy at the Raman frequency because excitation at $\omega_o$ cannot store energy in the conductor at $\omega_\ast$. Taken together these factors restrict SERS to the primary enhancement, which is dependent on $|\vec{E}|^2$, which is limited to factors of the order of $10^6$ to $10^7$ in the most favorable cases.

These results present a dilemma for the experimental observation of large scattering cross sections. Even adjacent inclusions would be unlikely to produce additional field amplifications of $10^7$–$10^8$ which would be required in order to achieve the reports of enhancement factors of $10^{14}$. The bandwidth restriction applies to any geometry, so geometries with very high quality factors have very narrowly peaked maximum gain regions. Moreover, large electric field gradients due to large curvatures can produce a considerable shift in the energy of the plasmon, which must be included in the theory for self-consistency. Thus, hot spots should shift the resonant energy significantly to the red in a very narrow bandwidth region. There is at present no direct experimental evidence for this phenomenon, which is required by the physics of the problem. An alternative view of the mechanism for enhancement includes the resonance Raman of the adsorbate molecule and the inclusion as one supermolecule. Our work suggests that holistic treatment of the problem as a resonance Raman phenomenon has a great deal of merit and should be systematically pursued.

Finally, we can postulate a situation where a dipole resonance occurs at one frequency and a higher multipole resonance at a different frequency. Because the radiation emerging from the passive radiator is due to the dipole term, any such higher multipole resonance must be associated with first enhancement. However, energy separations between the dipole and higher resonances are too large for this to be a viable explanation.

## VI. ACKNOWLEDGMENTS

DEA is pleased to acknowledge the hospitality of KyungHee University and associated support by the World Class University (WCU) program through the Korea Science and Engineering Foundation funded by the Ministry of Education, Science, and Technology under grant R33-2008-0000-10118-0. SF wishes to acknowledge the hospitality and associated support of Zhejiang University and the Guangbiao Professor program.